\pdfoutput=1 
\documentclass{JINST}
\let\ifpdf\relax

\title{Pressurised xenon as scintillator for gamma spectroscopy}

\author{F. Resnati$^a$ {\it for the MODES-SNM collaboration}\\
\llap{$^a$}ETH Zurich\\
IPP Schafmattstr. 20, 8093 Zurich, Switzerland\\
E-mail: \email{filippo.resnati@cern.com}}

\abstract{Detectors based on liquid or gas xenon have been used and are in use for a number of applications, in particular for the detection of gamma rays.
Xenon is a well-suited medium for gamma spectroscopy thanks to its high atomic number and, consequently, large cross-section for photo-electric absorption.
This paper presents experimental studies of high pressure xenon as a scintillator, with the aim of developing a gamma ray detector for the detection of Special Nuclear Materials (SNM).
The first goal was to study the dependence of the light yield and of the energy resolution on the thermodynamic conditions.
We present preliminary results from an optimised version of the detector.}


\begin{document}

\section{Introduction}
MOdular DEtection System for Special Nuclear Materials (MODES-SNM)~\cite{modes} is a FP7 funded project aiming at the design, construction, testing and qualification of a mobile system for the detection of Special Nuclear Materials (SNM), i.e.\ enriched uranium and plutonium.
The device is compatible with the IAEA's norms pertaining to Portable Radiation Scanner (PRS)~\cite{prs}, which is defined as a device designed mainly for covert detection of unauthorised or undeclared activities.
The final system will consist of an array of gamma and neutron (thermal and fast) detectors, which will improve the detection of radioactive sources in terms of sensitivity for shielded SNM.
The project aims at delivering a device ready to be operated by end-users.
The common technology for the MODES-SNM detector suite is given by high pressure cells filled with noble gases (xenon and helium) used as scintillators.
The $^4$He detectors for fast neutrons had already been developed by Arktis Radiation Detectors Ltd~\cite{arktis}, while the development of the xenon detector for gamma rays and of the helium detector for slow neutrons is part of the MODES-SNM project.
The cross-section of $^4$He for elastic scattering of fast neutron is peaked in the energy region between 1~MeV and 5~MeV, which matches the energy of fission neutrons.
The gamma ray cross-section is so small that a helium based detector is only mildly affected even by an intense gamma flux.

Xenon is an attractive material for gamma ray detection, mainly because of its high atomic number (Z = 54) and large cross-section for photo-electric absorption (photo-electric cross-section dominates up to around 300~keV).
Being a scintillation detector, the xenon spectrometer is not susceptible to vibrational disturbance and microphonic effects typically associated to ionisation detectors.
Unlike crystal scintillators, the detector is not prone to temperature-stress related fracturing, and below the supercritical point of xenon (15.8~$^\circ$C, 58~bar), the detector performance is largely unaffected by temperature variations.
Xenon scintillation has a fast decay time of about 40~ns, yielding an excellent time resolution.

In the following sections the latest results from the development of the gamma ray detector are presented.
For an overview of the MODES-SNM project, requirement and status we refer the reader to~\cite{curioni:2013}.

\section{Experimental apparatus}
\begin{figure}[tbp]
\centering
\includegraphics[width=.49\textwidth]{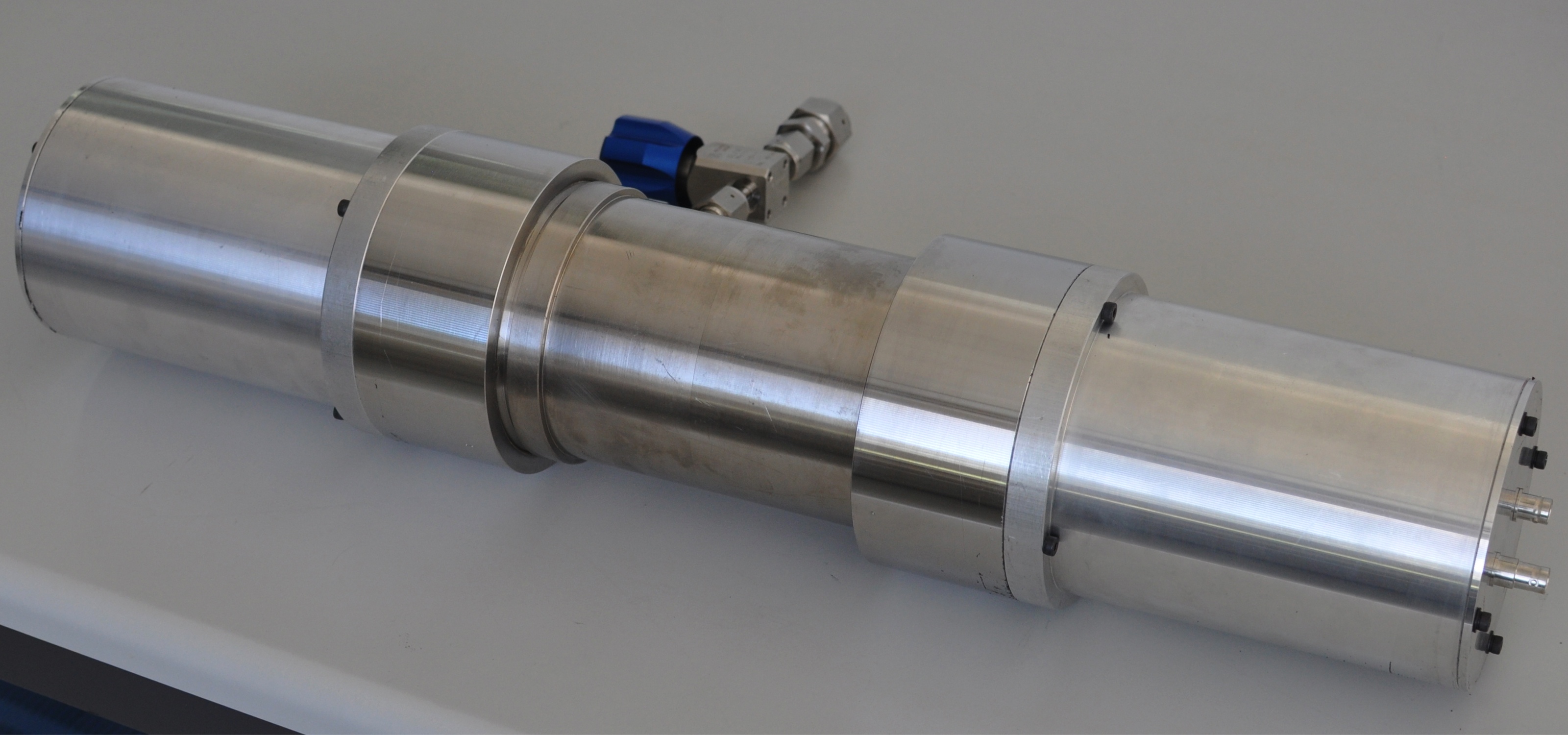}
\includegraphics[width=.49\textwidth]{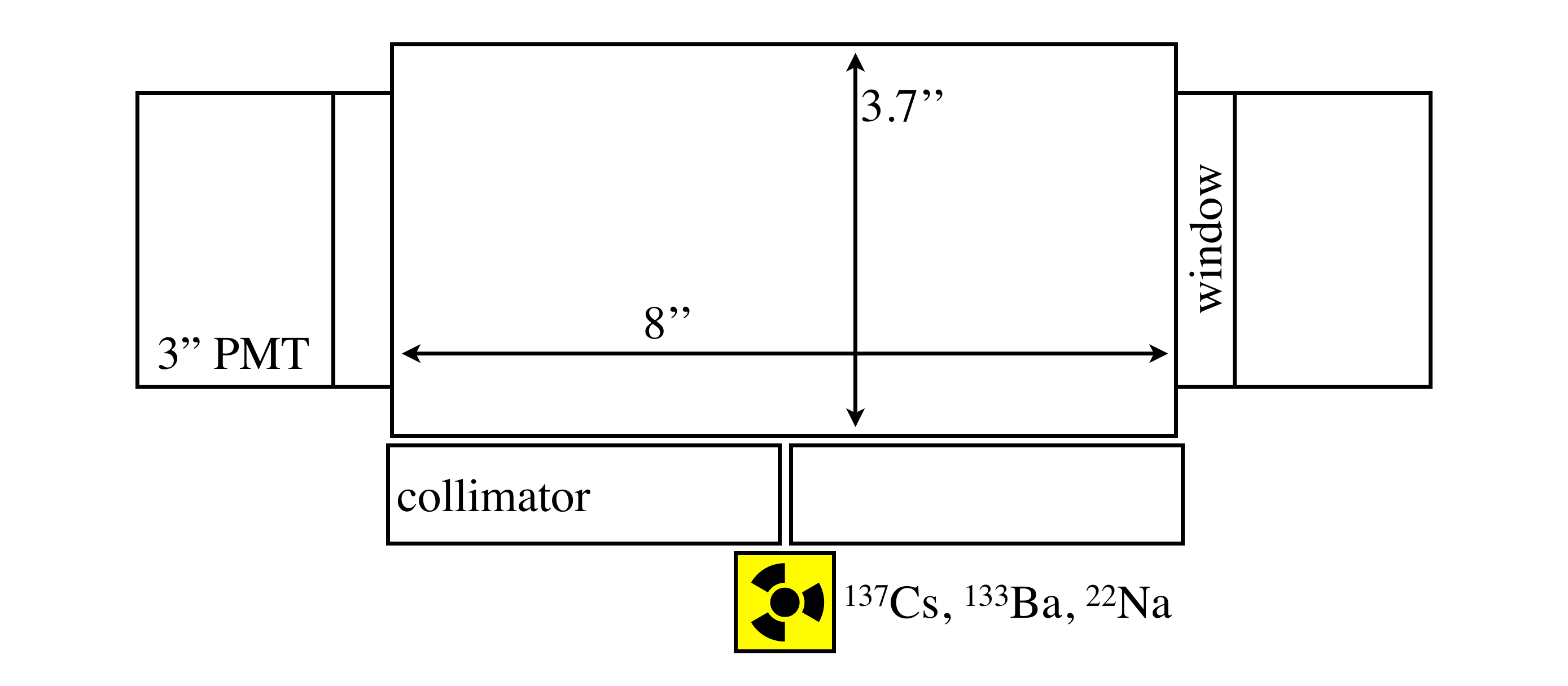}
\caption{Left: picture of the high pressure xenon tube. Right: schematic of the detector.}
\label{fig:detector}
\end{figure}

Results from the first development of the technology involving pressurised xenon at room temperature as scintillation medium for gamma spectroscopy are reported in~\cite{resnati:2013}.
At the time the detector was not optimised and suffered from significant systematic effects that worsened its performance.
We developed a new detector design that improves the light collection efficiency and reduces the passive material, yielding a better energy resolution and larger peak to total ratio (fraction of events in the full energy peak with respect to the totality of the events).
Fig.\ref{fig:detector} shows a picture and a schematic representation of the detector.
The xenon gas is contained in a certified titanium pressure vessel with a wall thickness of 2.5~mm.
The detector has a total weight of 8~kg, and the sensitive volume is a cylinder with a diameter of 9.4~cm and a length of 20.3~cm.
The scintillation light from xenon is peaked at 175~nm and is therefore rather difficult to be efficiently detected.
For this matter the inner surface of the tube is lined with a reflector combined with a wavelength shifter, that shifts the VUV light from xenon scintillation into visible light.
Two Hamamatsu R6233-100 high quantum efficiency Photo Multiplier Tubes (PMTs) readout the light through two transparent windows on opposite sides of the scintillation volume.
The proprietary construction (Arktis Radiation Detectors Ltd~\cite{arktis}) enables the sealed system to maintain high gas purity indefinitely, without requiring recirculation or re-purification of the xenon.
The two PMT waveforms are digitised independently with a CAEN V1751 ADC (1~GS/s, 10~bit).
The calibration of the PMTs is performed using the isolated single photo-electrons (p.e.) on the tail of the signals.
The integral of the signals is the quantity considered to evaluate the amount of detected light.
We took data with various gamma ray sources, collimated with lead bricks, having a 3~mm wide slit illuminating the detector.

\section{Results}
\begin{figure}[tbp]
\centering
\includegraphics[width=.49\textwidth]{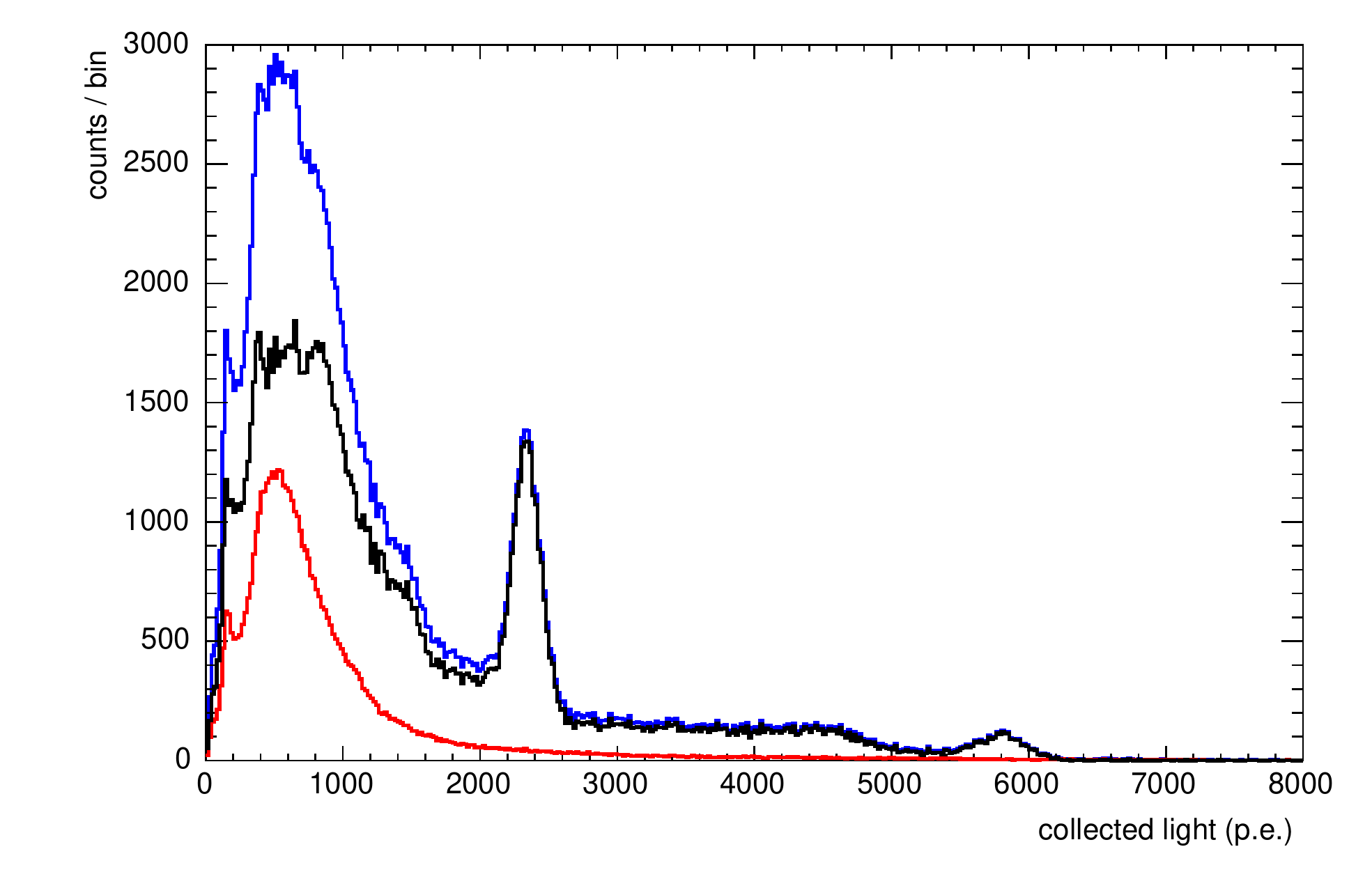}
\includegraphics[width=.49\textwidth]{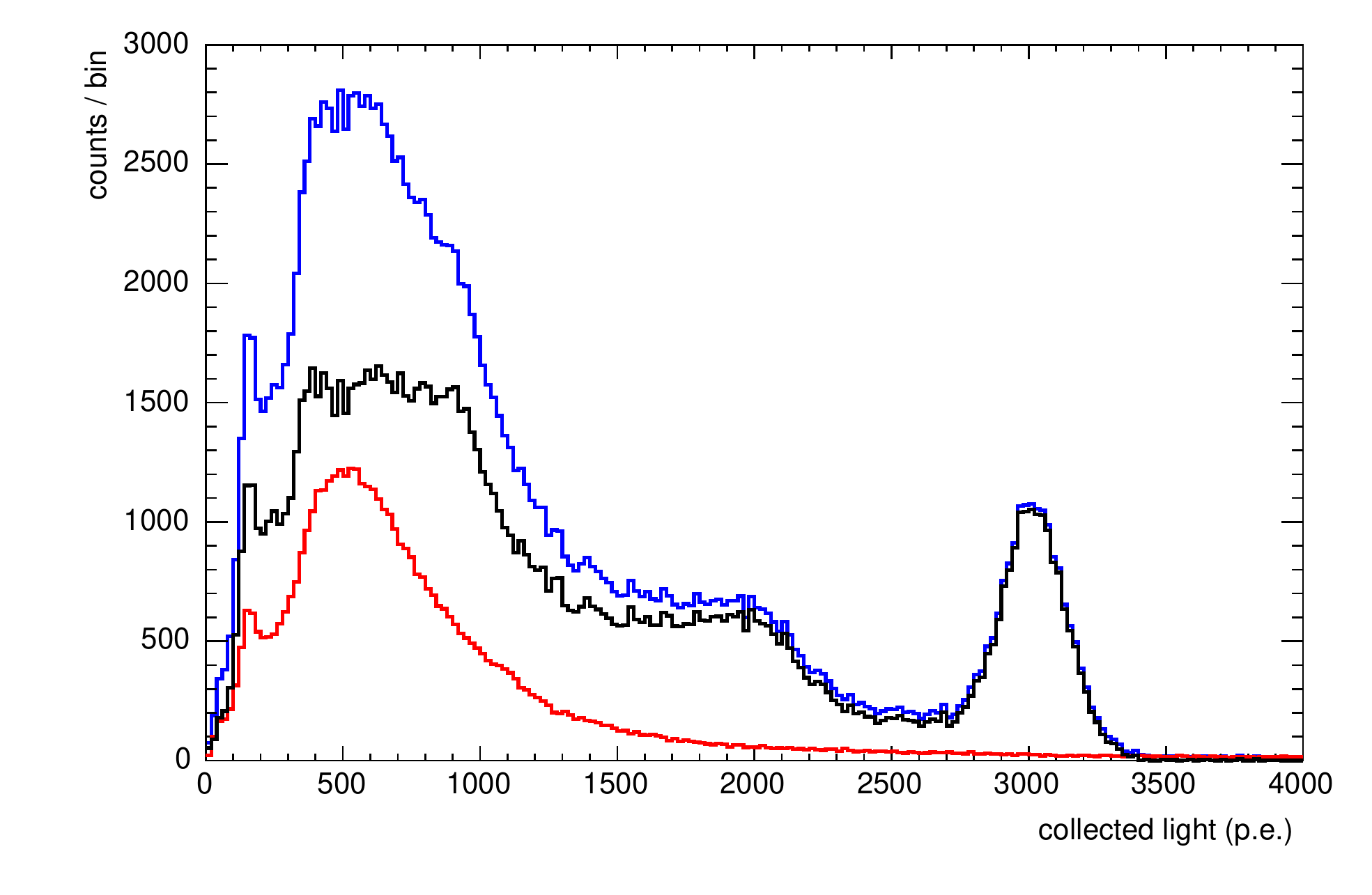}
\caption{Distribution of the collected light exposing the detector to $^{22}$Na (left) and $^{137}$Cs (right) sources. The histograms show the background in red, the sum of the signal and the background in blue and the signal only (background subtracted) in black.}
\label{fig:spectra}
\end{figure}

The detected light spectra (proportional to the energy released in the active volume) from exposing the detector (filled with 49~bar of pure xenon) to $^{22}$Na and $^{137}$Cs sources are shown in Fig.~\ref{fig:spectra}.
$^{22}$Na source provides lines at 511~keV and 1275~keV, while $^{137}$Cs source gives a line at 662~keV.
In this configuration, at 662~keV the light yield is 4.55~p.e./keV and the energy resolution (FWHM) is 8.6\%.
Comparing these results with the ones from the first detector prototype~\cite{resnati:2013}, we obtained a significant improvement of the light yield (formerly 1.9~p.e./keV), due to a better aspect ratio of the tube (formerly about 5:1) and a more efficient reflector.
Comparing the $^{137}$Cs spectra, one can appreciate the increase of the peak to total ratio, mainly due to the reduced thickness of the tube walls, and the decrease of the Compton edge, due to the larger active mass.

\begin{figure}[tbp]
\centering
\includegraphics[width=.49\textwidth]{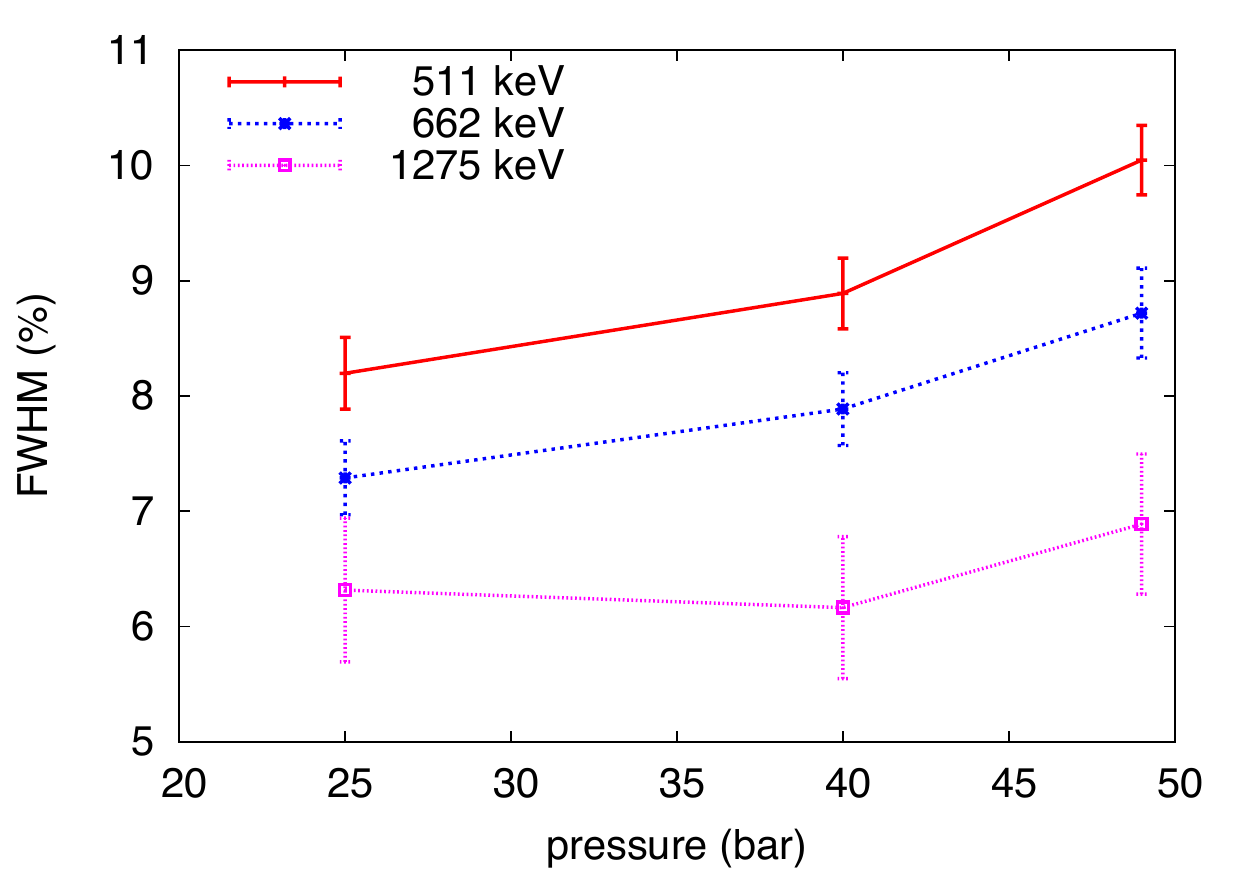}
\includegraphics[width=.49\textwidth]{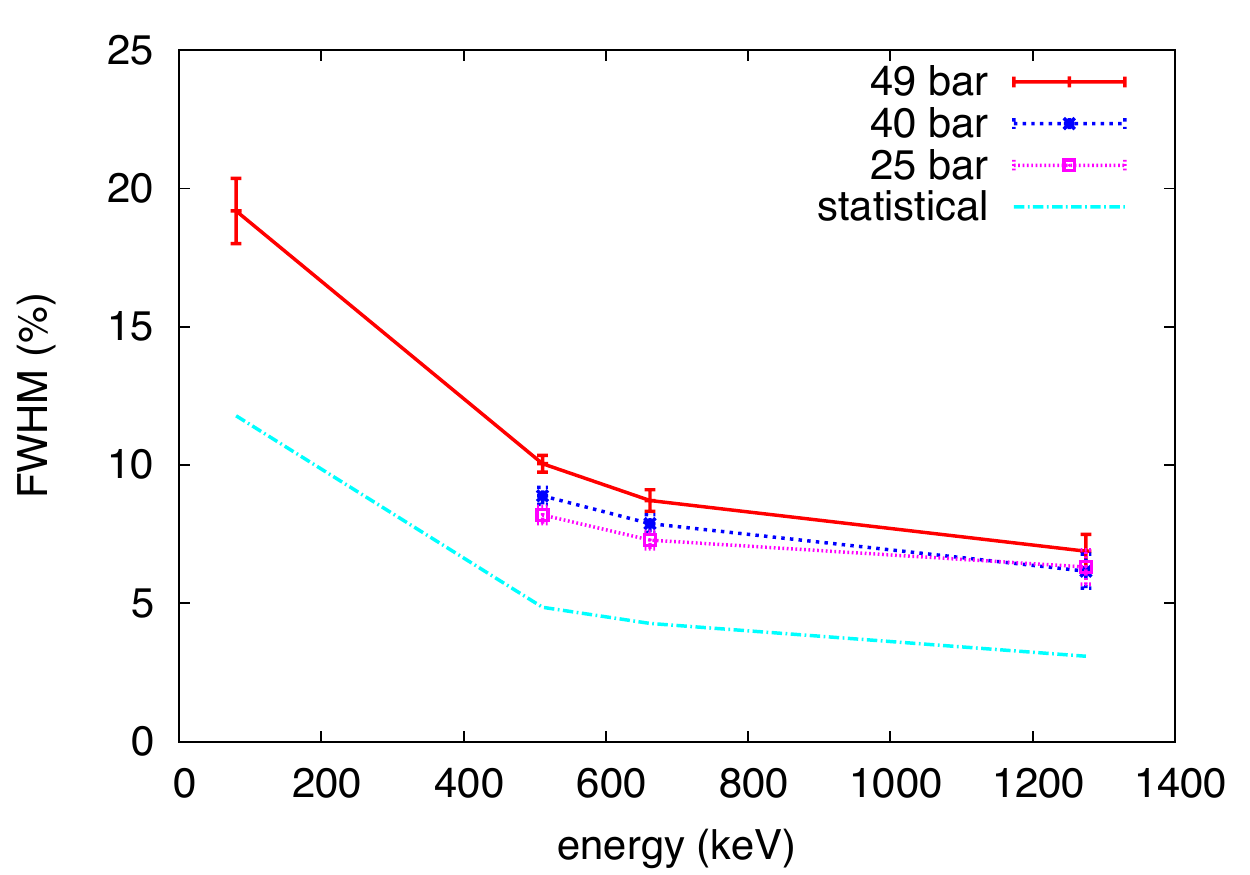}
\caption{Energy resolution as a function of pressure (left) and energy (right). A $^{133}$Ba source was used to provide the 80~keV line.}
\label{fig:FWHM}
\end{figure}

Fig.~\ref{fig:FWHM} (left) shows the energy resolution as a function of the pressure for three different gamma energies.
It is important to understand the dependence of the detector performance on the xenon pressure because an efficient detector should be operated at the largest possible density.
We clearly see that the energy resolution deteriorates when the xenon gas gets close to the supercritical phase (15.8~$^\circ$C, 58~bar), essentially because many scattering centres within the fluid hamper the uniform propagation of light.
The variation of the optical properties of the xenon is easily observable by simply watching inside the tube while changing the pressure: the xenon goes from a perfectly transparent medium to a turbid one.
For highly energetic gammas the containment in the active volume of the converted electron from Compton or photo-electric effects is not ensured at low densities, and this reflects in the worsening of the energy resolution of the 1275~keV line at 25~bar.

Fig.~\ref{fig:FWHM} (right) also shows, as a function of energy, the measured and expected energy resolution.
The latter is assuming that the spread is only due to statistical fluctuations.
The measured resolution improves with energy because of the photo-statistic contribution, but it is consistently worse than the expected one, pointing to a significant systematic effect.
This was found to be related to a non-uniform wavelength shifter coating on the reflector.
For the sake of the light collection uniformity, wavelength shifter coating is also foreseen on the transparent windows.

Finally, with a calibrated $^{137}$Cs source located 1.85~m away from the detector we measured a detection efficiency of 70\% at 49~bar, integrating over the full energy spectrum (threshold at 20~keV), which is comparable to the efficiency of a 2"x2" NaI crystal~\cite{stgobain}.

\section{Conclusions and outlooks}
We have reported experimental results that give an initial characterisation of the properties of high pressure xenon gas as a scintillator, in particular in terms of the energy resolution.
We find these results encouraging for applications of high pressure xenon scintillator detectors in gamma ray spectroscopy.
Presently the detector has a light yield at 662~keV of 4.55~p.e./keV.
The best energy resolution is 7.3\% (FWHM) obtained at 25~bar.
A detection efficiency of 70\% was measured at 49~bar.
These results confirm that, pending few improvements, such a technology can compete with typical scintillator crystal.
Compared to crystals the high pressure xenon tube is not fragile, which makes it ideal for applications where the detector must be movable, such as MODES-SNM.
Presently the detector meets the MODES-SNM requirements, but further development are foreseen to make it more competitive with standard gamma spectroscopy solutions.
Without modifying the design of the pressure vessel we expect to improve the uniformity of the light collection, yielding an energy resolution of 7--8\% (FWHM) at 662~keV with the source non-collimated.

\acknowledgments
This work is supported by the European Union through the MODES-SNM project (Call FP7-SEC-2011-1, ERC Grant agreement no. 284842).

\end{document}